\documentclass[prl,aps,twocolumn,superscriptaddress,showpacs]{revtex4}
\usepackage{graphicx,rotating,subfigure,amsmath,amsfonts,amssymb,delarray}
\renewcommand{\vec}[1]{\boldsymbol #1}

\def\12{\frac{1}{2}}

\begin{document}
\bibliographystyle{apsrev}

\title{Thermally activated Peierls dimerization in ferromagnetic spin chains}

\author{Jesko Sirker}
\email[]{j.sirker@fkf.mpg.de}
\affiliation{ Max-Planck-Institut f\"ur Festk\"orperforschung,
              Heisenbergstrasse 1, D-70569 Stuttgart, Germany }

\author{Alexander Herzog}
\affiliation{ Max-Planck-Institut f\"ur Festk\"orperforschung,
              Heisenbergstrasse 1, D-70569 Stuttgart, Germany }

\author{Andrzej M. Ole\'s}
\affiliation{ Max-Planck-Institut f\"ur Festk\"orperforschung,
              Heisenbergstrasse 1, D-70569 Stuttgart, Germany }
\affiliation{ Marian Smoluchowski Institute of Physics, Jagellonian
              University, Reymonta 4, PL-30059 Krak\'ow, Poland }

\author{Peter Horsch}
\affiliation{ Max-Planck-Institut f\"ur Festk\"orperforschung,
              Heisenbergstrasse 1, D-70569 Stuttgart, Germany }

\date{\today}

\begin{abstract}
We demonstrate that a Peierls dimerization can 
occur in ferromagnetic spin chains activated by thermal fluctuations. 
The dimer order parameter and entanglement measures are studied as 
functions of the modulation of the magnetic exchange interaction and 
temperature, using a spin--wave theory and the density--matrix 
renormalization group. We discuss the case where a periodic modulation 
is caused by spin--phonon coupling and the case where electronic
states effectively induce such a modulation. The importance of the 
latter for a number of transition metal oxides is highlighted.
\end{abstract}


\maketitle

Structural instabilities of electronic systems can occur due to the coupling
of electronic and lattice degrees of freedom (phonons). They are particularly
important for quasi one--dimensional (1D) systems where the gain in electronic
energy due to a lattice distortion often outweighs the cost in elastic energy.
A well known example is the Peierls instability \cite{Peierls} of the 1D free
electron system towards a static lattice distortion determined by the Fermi
momentum.  For a commensurate distortion, an excitation gap is opened turning
a metallic system into a band insulator. This Peierls metal--insulator
transition plays an important role, for example, in organic charge--transfer
solids \cite{Gruner}. A related instability occurs for antiferromagnetic (AFM)
spin chains coupled to phonons.  Here magnetic energy is gained by distorting
the lattice which can lead to the so called spin--Peierls (SP) transition.
Although a SP phase transition was first observed in organic materials
\cite{BrayHart}, it was the discovery of such a transition in CuGeO$_3$ by
Hase {\it et al.}  \cite{Hase} that has led to great interest in these
phenomena \cite{Joh00}.

Quite recently, another type of Peierls instability for spin chains has 
been found which is not driven by spin--phonon coupling but rather by a 
coupling of the spins with electronic degrees of freedom (orbitals). 
Here a ferromagnetic (FM) spin chain shows a periodic modulation 
(dimerization) of the magnetic exchange in a certain finite temperature 
region while the ground state is the uniform fully polarized FM state
\cite{SirkerKhaliullin}. In \cite{KeimerSirker, HorschKhaliullin} it 
has been argued that this mechanism is responsible for the remarkable 
physical properties of YVO$_3$ in the finite temperature C--type AFM 
phase. Clearly, the lattice will react to a modulation of the magnetic 
exchange, however, spin--phonon coupling is not the driving force in 
this case and any lattice distortion is only a secondary effect.

In this Letter we want to establish general mechanisms which can drive 
a Peierls dimerization in FM spin chains. To highlight the differences 
between AFM and FM chains we will first consider a coupling to lattice 
degrees of freedom. The phonons are often treated adiabatically which
is justified if the phonon frequency is smaller than the Peierls gap. 
In the adiabatic approximation the Hamiltonian can be written as 
$H=H_{\rm mag}+E_{\rm el}$ with
\begin{equation}
\label{eq1}
H_{\rm mag}=J\sum_{j=1}^N \left\{1+(-1)^j\delta\right\}\vec{S}_j\!\cdot\!\vec{S}_{j+1}\;,
\end{equation}
and $E_{\rm el}=NK\delta^2/2$. Here $J$ is the exchange constant, 
$\vec{S}_j$ is a spin $S$ operator at site $j$, and $N$ is the number 
of sites. In the absence of a magnetic field the modulation is expected 
to be commensurate with wave vector $k=\pi$ and is parameterized by 
$\delta\in [0,1]$. The elastic energy $E_{\rm el}$ contains the elastic 
constant $K$. Both parameters $K,\,\delta$ can be related to the 
spin--phonon interaction strength. Note that writing the Hamiltonian as 
a sum of a magnetic and an elastic part as in Eq.~(\ref{eq1}) 
corresponds to the random--phase approximation (RPA) by Cross and Fisher 
\cite{CrossFisher}. Although the model (\ref{eq1}) is strictly 1D, the 
static, mean--field (MF) treatment of the three--dimensional phonons 
allows for a finite temperature phase transition if $\delta(T)$ is
treated as a thermodynamical degree of freedom determined by minimizing 
the free energy.

Let us start with the case where $\vec{S}_j\!\cdot\!\vec{S}_{j+1}\to
S_j^xS_{j+1}^x+S_j^yS_{j+1}^y=(S^+_jS^-_{j+1} + S^-_{j}S^+_{j+1})/2$,
i.e., we replace the SU(2)--symmetric spin exchange by an $XX$-type of
interaction. In this case the sign of $J$ does not matter and the
system becomes equivalent to a free spinless fermion model by
Jordan--Wigner transformation. The Hamiltonian is then easily
diagonalized by Fourier transformation and in the ground state for
small $\delta$ one finds a gain in magnetic energy $E_{\rm
  mag}\sim\delta^2\ln\delta$. This outweighs the cost in elastic
energy $E_{\rm el}\sim\delta^2$ and constitutes the Peierls
instability for lattice fermions \cite{Pincus}. For the isotropic
antiferromagnet ($J>0$, SU(2)--symmetric exchange), field theoretical
arguments show that $E_{\rm mag}\sim -\delta^{4/3}$
\cite{CrossFisher}. Again this outweighs the cost in elastic energy
leading to a SP transition and the opening of a spin gap,
$\Delta\sim\delta^{2/3}$.

Contrary to the two cases discussed above there is no gain in magnetic 
energy in the ground state for FM coupling, $J<0$. For $\delta\in [0,1)$ 
the ground state is always the fully polarized FM state. We will show in 
the following that thermal fluctuations can, however, activate a Peierls 
dimerization. We will use the density--matrix renormalization group 
applied to transfer matrices (TMRG) to study this effect. The TMRG 
algorithm is based on a mapping of the 1D quantum onto a two--dimensional 
classical system. A transfer matrix is then defined allowing it to perform 
the thermodynamic limit exactly, i.e., all the numerical results presented 
here will be directly for the infinite system. Details of the method can 
be found in \cite{Peschel,GlockeKluemperSirker_Rev,SirkerKluemperEPL}. 
In addition, we will also apply Takahashi's modified spin--wave theory 
(MSWT) \cite{TakahashiMSWT4} to this problem.

\begin{figure}[t!]
\includegraphics*[width=8.2cm]{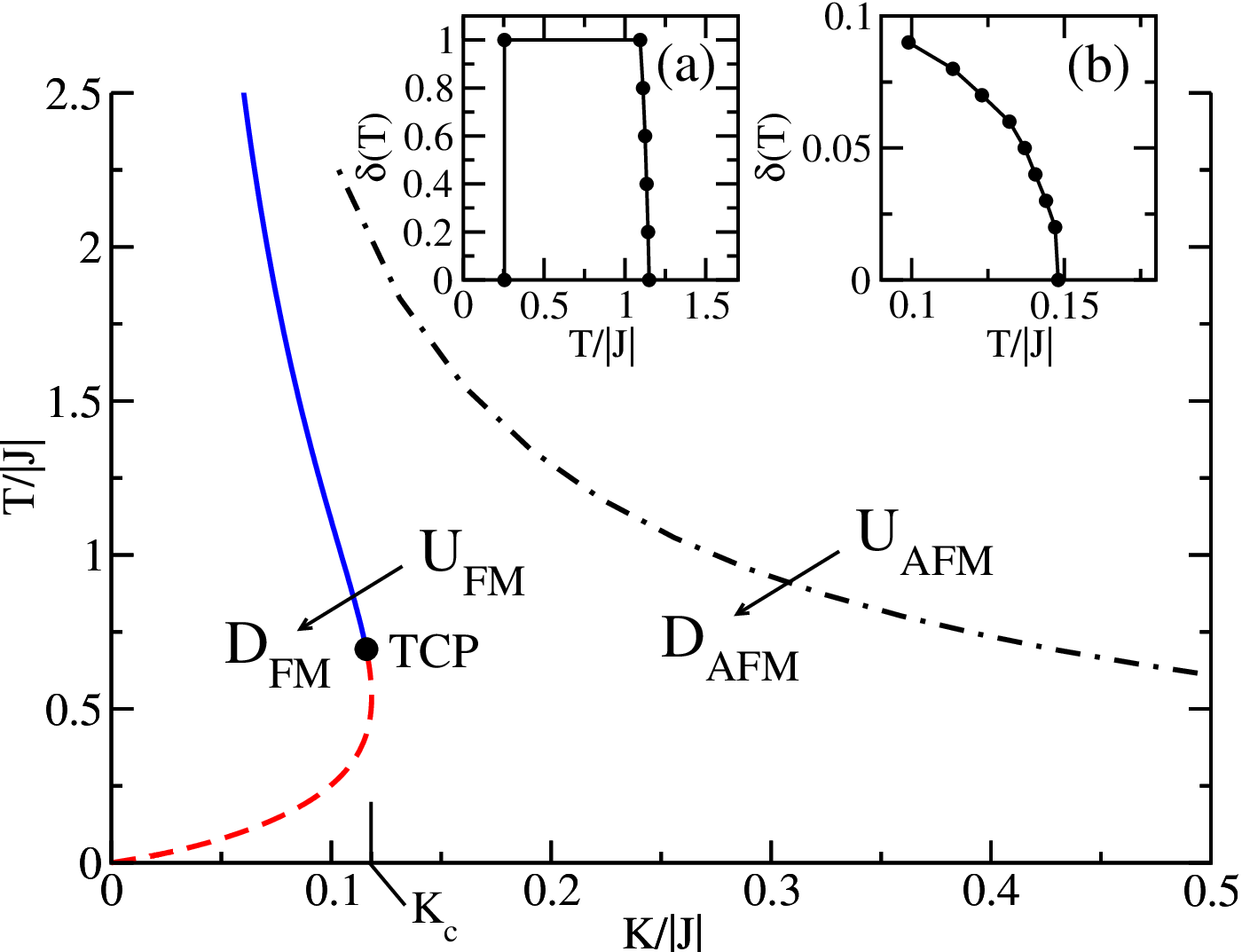}
\caption{(color online) 
Phase diagrams for the dimerized AFM and FM $S=1/2$ Heisenberg chains.  
The dot--dashed line depicts the second order SP transition from the 
uniform (U) to the dimerized (D) phase for the AFM chain. For the FM chain, 
the D phase exists only at finite $T$ (in units of $k_B=1$) and only if 
$K<K_c\simeq 0.118\,|J|$ --- the transition is either second or first
order, as shown by solid and dashed lines, respectively, 
and changes its character at the tricritical point (TCP).
The insets show the order parameter $\delta(T)$ for:
(a) the FM chain with $K/|J|=0.1$, and
(b) the AFM chain with $K/J=2$ 
(the lines are guides to the eye).}
\label{fig1}
\end{figure}

In Fig.~\ref{fig1} the phase diagrams for the $S=1/2$ isotropic AFM and FM
Heisenberg models as defined in Eq.~(\ref{eq1}) are shown. The phase
boundaries and order parameters are obtained using the TMRG algorithm. For the
AFM we have a dimerized phase for any value of the elastic constant $K/|J|$ at
low enough temperatures because the gain in magnetic energy will always win.
The phase transition is second order and the evolution of the order parameter
is exemplified for $K/|J|=2$ in inset (b) of Fig.~\ref{fig1}. For the FM, on
the other hand, a dimerized phase exists only at finite temperatures and only
if $K/|J| < K_c/|J|\simeq 0.118$. Here we find a tricritical point (TCP) at
$(T_{\rm TCP}/|J|,K_{\rm TCP}/|J|)\simeq (0.696,0.116)$. For $K<K_{\rm TCP}$
the transition is first (second) order if $T<T_{\rm TCP}$ ($T>T_{\rm TCP}$),
respectively. Inset (a) of Fig.~\ref{fig1} shows that the order parameter for
$K/|J|=0.1$ evolves indeed continuously at the upper phase boundary although
it increases very steeply to one.  Note that in the small window 
$K_{\rm TCP}<K<K_c$ both the high and the low temperature transition 
will be first order.

Next, we discuss the application of Takahashi's MSWT to this problem.  Usual
spin--wave theory is modified by introducing a Lagrange multiplier which
enforces a nonmagnetic state at finite temperature. This guarantees that the
Mermin--Wagner theorem is respected. For the isotropic FM chain, results
obtained by MSWT have been shown to be in excellent agreement at low
temperatures with exact results obtained by the Bethe ansatz
\cite{TakahashiMSWT,TakahashiMSWT4}. For the dimerized chain the unit cell is
doubled so that a Holstein--Primakoff transformation with different bosonic
operators on the two sublattices is required.  The diagonalized Hamiltonian in
linear spin--wave theory is then given by $H_{\rm
  mag}=Ne_0+\sum_k\big\{\omega_k^+\beta_k^\dagger\beta_k^{}
+\omega_k^-\alpha_k^\dagger\alpha^{}_k\big\}$, with $e_0=JS^2$ and the two
magnon branches
$\omega_k^\pm=2|J|S\left(1\pm\sqrt{\cos^2k+\delta^2\sin^2k}\right)$.  The
constraint of zero magnetization
$NS=\sum_k\left\{n_{B}(\omega_k^-)+n_{B}(\omega_k^+)\right\}$ is implemented
by a Lagrange multiplier $\mu$ which acts as a chemical potential with
$n_B(\omega_k^\pm)=\{\exp[(\omega_k^\pm-\mu)/T]-1\}^{-1}$ being the Bose
factors. For $t/(1-\delta^2)\ll 1$, where $t=T/(|J|S)$ is the reduced
temperature, we find analytically
$4S^2\mu/T=-t/(1-\delta^2)+\mathcal{O}([t/(1-\delta^2)]^{3/2})$.  In the same
limit the free energy per site is given by $(f-e_0)/T=\alpha
[t/(1-\delta^2)]^{1/2}+\mathcal{O}([t/(1-\delta^2)])$, with
$\alpha=-\zeta(3/2)/(2\sqrt{\pi})$. For the FM chain we have therefore a gain
in magnetic energy due to a dimerization $\sim -T^{3/2}\delta^2$.

\begin{figure}[t!]
\includegraphics*[width=8.2cm]{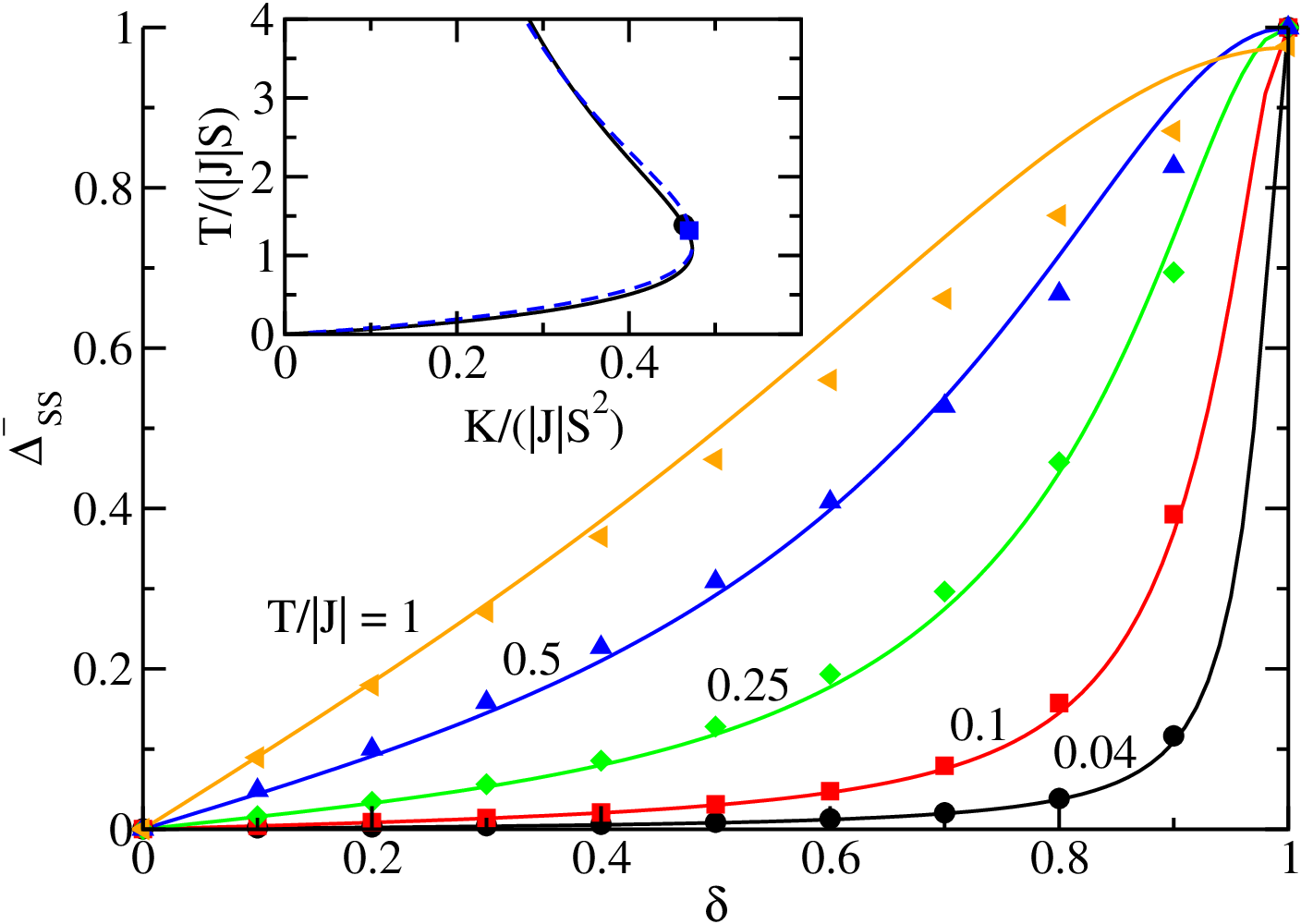}
\caption{(color online) Order parameter $\Delta^-_{SS}$ (\ref{Delta}) as a
  function of $\delta$ for the $S=1$ dimerized FM. The lines (symbols) denote
  the MSWT (TMRG) results, respectively. The inset shows the phase diagrams
  obtained by TMRG for the Hamiltonian (\ref{eq1}) with $S=1/2$ (solid line)
  and $S=1$ (dashed line) with both axes scaled appropriately.  The phase
  transition is first (second) order for $T<T_{\rm TCP}$ ($T>T_{\rm TCP}$).
  The TCP is marked by a dot (square) for $S=1/2$ ($S=1$).}
\label{fig2}
\end{figure}

To calculate spin correlation functions it is essential to take also 
quartic bosonic terms into account. For the bond correlations
$B_{s(w)}\equiv\langle\vec{S}_{2j}\!\cdot\!\vec{S}_{2j\pm 1}\rangle$ 
this leads to
\begin{equation}
  \label{SWT3}
  B_{s(w)}=\Big(\frac{1}{N}\sum_k
  \left\{n_B(\omega_k^-)-n_B(\omega_k^+)\right\}f^\pm_k\Big)^2,
\end{equation}
with $f^\pm_k=(\cos^2k\pm\delta\sin^2k)/\sqrt{\cos^2k+\delta^2\sin^2k}$. The
plus (minus) sign in $\langle\vec{S}_{2j}\!\cdot\!\vec{S}_{2j\pm 1}\rangle$
and in $f^\pm_k$ applies for the strong (weak) bond, respectively. We define
\begin{equation}
  \label{Delta}
  \Delta^\pm_{SS}= 
  \langle \vec{S}_{2j}\!\cdot\!\vec{S}_{2j+1}\rangle\pm\langle
  \vec{S}_{2j}\!\cdot\!\vec{S}_{2j-1}\rangle\,, 
\end{equation}
with $\Delta^-_{SS}$ acting as an order parameter for the dimerized chain. In
Fig.~\ref{fig2} the MSWT and TMRG results for $\Delta^-_{SS}$ are compared for
the case of $S=1$. The agreement is good for temperatures up to $T/|J|\sim 1$,
in particular for small $\delta$. We also note that the MSWT gives a
value in the fully dimerized case ($\delta=1$) which is in good agreement with
the exact result, however, it predicts
corrections for $\delta=1-\epsilon$ ($\epsilon\ll 1$) to be of order 
$\epsilon^2$, whereas the numerical results and perturbation theory show 
that the corrections are of order $\epsilon$. 
In the inset of Fig.~\ref{fig2} 
it is shown that the phase diagrams for model (\ref{eq1}) with $S=1/2$ 
and $S=1$ are almost identical, if the axes are scaled appropriately.

\begin{figure}[t!]
\includegraphics*[width=8.2cm]{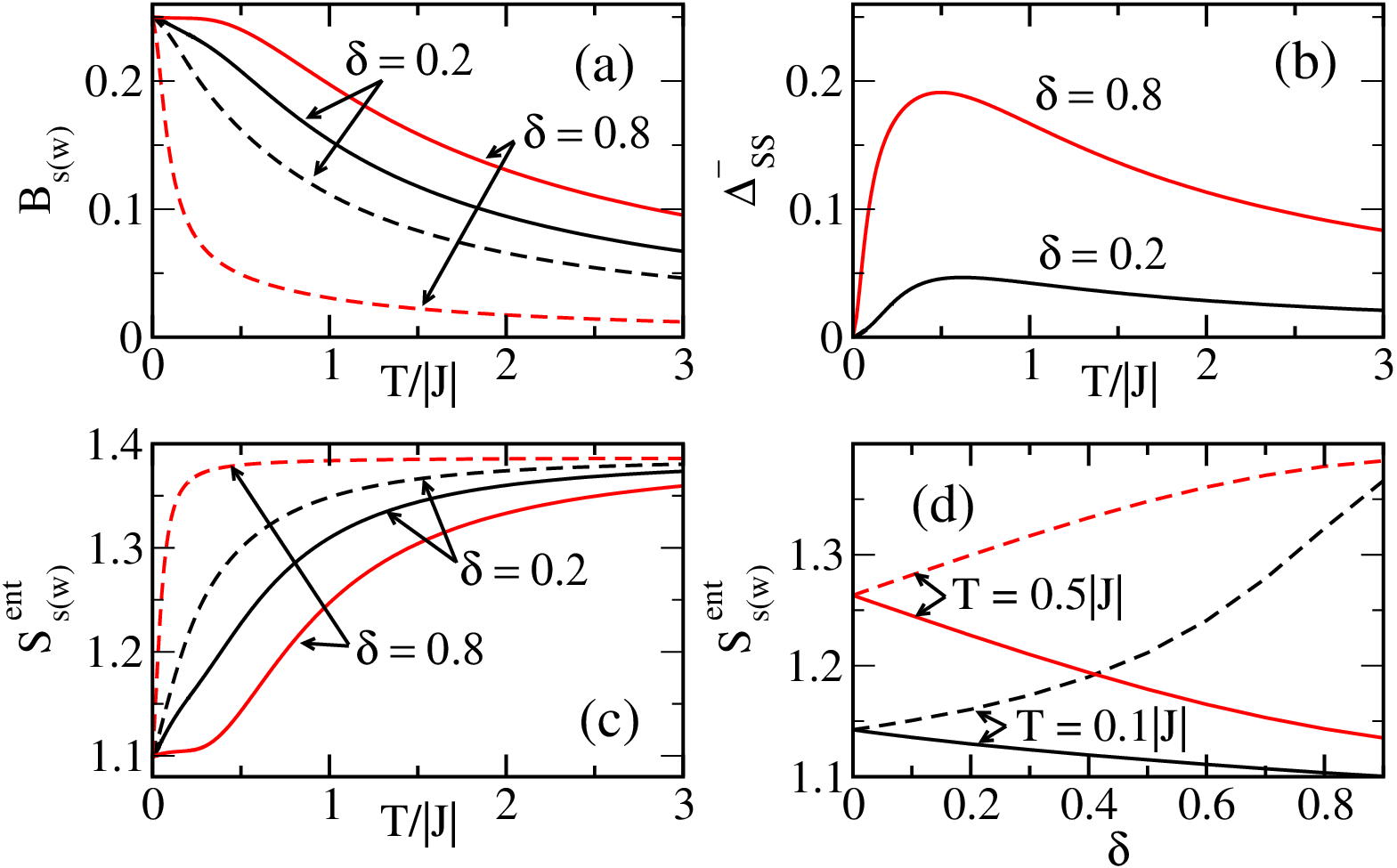}
\caption{(color online) 
TMRG results for $S=1/2$: 
(a) $B_{s(w)}$ for the strong (solid line) and weak bond (dashed line), 
(b) $\Delta^-_{SS}$ for the same values of $\delta$ as in (a),
(c) the entanglement entropy $S^{\rm\, ent}_{s(w)}$ for different $\delta$ 
    as a function of temperature, and 
(d) $S^{\rm\, ent}_{s(w)}$ as a function of $\delta$ for two temperatures.}
\label{fig3}
\end{figure}

In Fig.~\ref{fig3}(a) the correlation functions on the strong and weak bond
for $S=1/2$ are shown separately as a function of temperature for different
$\delta$.  We want to emphasize again that for $\delta\in [0,1)$ the ground
state is still the usual FM state and the correlations on the weak and strong
bond are thus identical, $B_s=B_w=1/4$.  The difference between the
correlations on the strong and on the weak bond, $\Delta^-_{SS}$, shown in
Fig.~\ref{fig3}(b) is therefore zero at $T=0$, goes through a maximum at some
finite temperature, and goes to zero again for $T\to\infty$ where $B_{s(w)}\to
0$.

Another way of looking at the response of the FM chain to a periodic
modulation is to study the {\em entanglement\/} of a weak or a strong bond 
with the rest of the system. Here we will concentrate on the case $S=1/2$. 
The entries of the two--qubit reduced density matrix $\tilde{\rho}$ for a 
bond can be related to the correlation functions on that bond 
\cite{GlaserBuettner,Pratt}.
The concurrence for $\tilde\rho$ --- an entanglement measure
commonly used at zero temperature --- can be expressed as $C_{2j,2j\pm
  1}(\tilde\rho) = 2\max\{0,|\langle S^+_{2j}S^-_{2j\pm 1} \rangle | -
| 1/4 + \langle S^z_{2j}S^z_{2j\pm 1}\rangle |\}$. It is zero for
FM correlations. More interesting is the behavior of the
entanglement entropy, ${\cal S}^{\rm\, ent}_{s(w)} =
-\mbox{Tr}\,\tilde\rho\ln\tilde\rho$.  It is again zero for the fully
polarized ground state which is a pure state.  At finite temperature we
have for $\alpha={s(w)}$
\begin{equation}
\label{Sent}
{\cal S}^{\rm\, ent}_{\alpha}\!=\!\Big(B_{\alpha}-\frac{1}{4}\Big)\ln\Big(\frac{1}{4}-B_{\alpha}\Big)
-\!\Big(B_{\alpha}+\frac{3}{4}\Big)\ln\Big(\frac{1}{4}+\frac{B_{\alpha}}{3}\Big)\, .
\end{equation}
For $T\to 0$, $B_{s(w)}\to 1/4$ and ${\cal S}^{\rm\, ent}_{s(w)}\to\ln 3$, see
Fig.~\ref{fig3}(c).  ${\cal S}^{\rm\, ent}_{s(w)}$ therefore jumps signaling
the phase transition at $T=0$. For $T\to\infty$, on the other hand,
$B_{s(w)}\to 0$ and ${\cal S}^{\rm\, ent}_{s(w)}\to 2\ln 2$. Quite generally,
the entanglement entropy for a segment with $n$ sites will go to $n{\cal S}_T$
for $T\to\infty$, where ${\cal S}_{T}$ is the thermal entropy per site
\cite{SoerensenAffleck}. At any fixed finite temperature the entanglement
entropy ${\cal S}^{\rm\, ent}_{s(w)}$ decreases (increases) on the strong
(weak) bond with increasing modulation $\delta$, see Fig.~\ref{fig3}(d).  The
gain in magnetic energy at finite temperature due to a dimerization might
therefore also be seen as a gain in entanglement entropy on the weak bonds.

\begin{figure}[t!]
\includegraphics*[width=8cm]{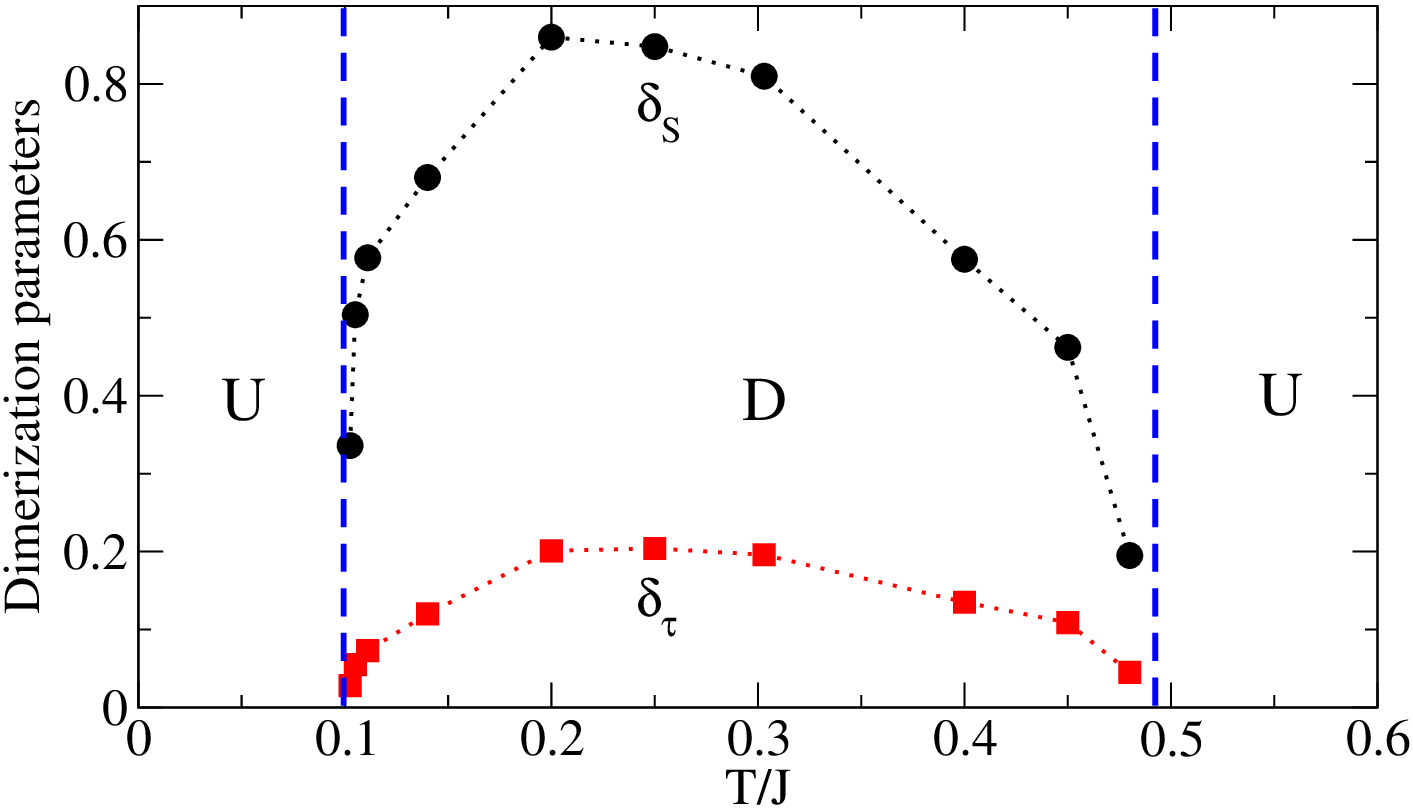}
\caption{(color online) Phase diagram and dimerization parameters 
$\delta_S,\,\delta_\tau$ for the spin--orbital model (\ref{SO1}) with 
$\gamma_H=0.1$ in MF decoupling. The dashed lines denote the 
phase boundaries between the uniform (U) and dimerized (D) phases.}
\label{fig4}
\end{figure}

Let us finally discuss the relevance of a thermally driven dimerization for
systems with orbital degrees of freedom. This mechanism is particularly
important for transition metal oxides with perovskite structure where the
valence electrons are situated in the $t_{2g}$ orbitals. Because
$t_{2g}$ orbitals are not bond oriented the electron--phonon coupling is weak
so that we might ignore lattice degrees of freedom to first approximation. 
With appropriately rescaled parameters, the physics discussed below is almost
independent of the spin value $S$. For definiteness, we will consider in the
following the case of an effective spin $S=1$ appropriate for systems with a
$3d^2$ valence electron configuration, as for example, YVO$_3$, and a twofold
orbital degeneracy described by an orbital pseudospin $\tau=1/2$. A 1D
Hamiltonian reflecting the spin--orbital physics for such a system is given by
\cite{KhaliullinHorsch}
 \begin{equation}
\label{SO1}
H_{S\tau}=J\sum_j\left(\vec{S}_j\!\cdot\!\vec{S}_{j+1}+1\right)
\left(\vec{\tau}_j\!\cdot\!\vec{\tau}_{j+1}+\frac{1}{4}-\gamma_H\right),
\end{equation} 
where $J>0$ is the superexchange and $\gamma_H$ is proportional to the 
Hund's coupling and promotes FM spin correlations. 
Using a MF decoupling, which is reasonable for FM spin correlations
\cite{Ole06}, we write $H_{S\tau}\simeq H_S+H_\tau$, where $H_{S}$
($H_{\tau}$) is the Hamiltonian for the spin (orbital) sector, respectively.
If we allow for a dimerization in both sectors then $H_{S(\tau)}$ is --- up to
a constant --- given by Eq. (\ref{eq1}) with $J\to J\mathcal{J}_{S(\tau)}$,
$\delta\to\delta_{S(\tau)}$, and $\vec{S}$ representing the spins $S=1$ or the
orbital pseudospins $\tau=1/2$, respectively. The effective superexchange
constants are given by $\mathcal{J}_S = \Delta_{\tau\tau}^+/2
+1/4-\gamma_H$ and $\mathcal{J}_\tau = \Delta^+_{SS}/2 +1$, with
$\Delta_{\tau\tau}^\pm$ defined analogously to $\Delta^\pm_{SS}$. 
Strong quantum fluctuations for pseudospin $1/2$ and $\gamma_H>0$ will
favor AFM coupled orbitals, $\mathcal{J}_\tau > 0$, and FM coupled
spins, $\mathcal{J}_S < 0$.  The dimerizations are then given by
$\delta_S = \Delta^-_{\tau\tau}/(2\mathcal{J}_S)$ and $\delta_\tau =
\Delta^-_{SS}/(2\mathcal{J}_\tau)$. This means that the exchange
constants and the dimerizations for each sector are determined by the
nearest--neighbor correlations in the other sector and therefore have
to be calculated self--consistently. We can simplify this procedure by
noting that
$\Delta^+_{SS(\tau\tau)}$ show only a weak dependence on dimerization and
temperature for low temperatures. We therefore fix $\mathcal{J}_{S(\tau)}$ by
using the values for $\Delta^+_{SS(\tau\tau)}$ obtained for an undimerized
chain at zero temperature.  This leads to $\mathcal{J}_\tau = 2$ and
$\mathcal{J}_S = 1/2-\ln 2-\gamma_H$ \footnote{Intersite correlations for the
  undimerized orbital $\tau=1/2$ chain with $\mathcal{J}_{\tau}>0$ are
  $\langle\vec{\tau}_j\!\cdot\!\vec{\tau}_{j+1}\rangle = 1/4-\ln 2$.}.
Now the dimerizations $\delta_{S(\tau)}$ can be easily
determined self--consistently. The results for $\gamma_H=0.1$ --- which is a
realistic value for cubic vanadates --- are shown in Fig.~\ref{fig4}.
For $0.10\lesssim T/J\lesssim 0.49$ the self--consistent MF decoupling 
leads to nonzero values for $\delta_{S(\tau)}$. The evolution 
of the dimerization parameters in this temperature regime has a 
dome--shaped form with a maximum at $T/J\sim 0.2$. In agreement with 
Fig.~\ref{fig1}, the dimerization in the FM spin chain is much
larger than the dimerization in the AFM orbital chain and at $T/J= 0.2$ we
have $\delta_S\approx 0.86$ which is already close to perfect dimerization
(Fig.~\ref{fig4}).  This underlines that the thermally activated dimerization
in the FM chain is the driving force behind the finite temperature dimerized
phase for the spin--orbital chain. The phase transitions at finite temperature
between a uniform and a dimerized phase are a consequence of the MF
decoupling.  Such phase transitions will not occur for the strictly 1D model
(\ref{SO1}).  Nevertheless, numerical calculations for this model
\cite{SirkerKhaliullin} show that a dimerization is the leading instability at
temperatures which support the dimerized phase in the MF decoupling solution.

Summarizing, we have shown that a dimerization can occur in FM spin chains but
has to be activated by thermal fluctuations. The gain in magnetic energy at
finite temperatures can be related to an increased entanglement entropy on the
weak bonds. For a FM chain with spin--phonon coupling we have derived the
phase diagrams as a function of temperature $T$ and the effective elastic
constant $K$ for spin values $S=1/2$ and $S=1$. Thermodynamic properties of
the dimerized FM chain can be calculated analytically with good accuracy for
temperatures $T\lesssim |J|S$ by a MSWT. Remarkably, this approach works for
all dimerizations $\delta\in [0,1]$ if quartic terms are taken into account
appropriately.  For a system of coupled FM spin-$1$ and AFM orbital
pseudospin-$1/2$ degrees of freedom we found, using a mean--field decoupling,
a finite temperature dimerized phase. This shows that a dimerization is a
universal instability of FM chains at finite temperatures, and may be
triggered by the coupling to purely electronic degrees of freedom.  This
latter mechanism seems to be relevant for many transition metal oxides with
(nearly) degenerate orbital states.

The authors thank G.~Khaliullin for valuable discussions.  A.M.~Ole\'s
acknowledges support by the Foundation for Polish Science (FNP) and by the
Polish Ministry of Science and Education Project No.~N202 068 32/1481.


\end{document}